\documentclass[preprint,12pt,authoryear,appendixfloats]{aastex61}

\usepackage{float}
\usepackage{lineno}
\begin{document}

\title{Tropospheric Composition and Circulation of Uranus with ALMA and the VLA}

\author{Edward M. Molter}
\affiliation{Astronomy Department, University of California, Berkeley; Berkeley CA, 94720, USA}
\correspondingauthor{Edward Molter}
\email{emolter@berkeley.edu}

\author{Imke de Pater}
\affiliation{Astronomy Department, University of California, Berkeley; Berkeley CA, 94720, USA}
\affiliation{Faculty of Aerospace Engineering, Delft University of Technology, Delft 2629 HS, The Netherlands}

\author{Statia Luszcz-Cook}
\affiliation{Department of Astronomy, Columbia University, Pupin Hall, 538 West 120th Street, New York City, NY 10027}
\affiliation{Astrophysics Department, American Museum of Natural History, Central Park West at 79th Street, New York, NY 10024, USA}

\author{Joshua Tollefson}
\affiliation{Astronomy Department, University of California, Berkeley; Berkeley CA, 94720, USA}

\author{Robert J. Sault}
\affiliation{School of Physics, University of Melbourne, Victoria, Australia}

\author{Bryan Butler}
\affiliation{National Radio Astronomy Observatory, Socorro, NM, United States}

\author{David de Boer}
\affiliation{Astronomy Department, University of California, Berkeley; Berkeley CA, 94720, USA}

\begin{abstract}

We present ALMA and VLA spatial maps of the Uranian atmosphere taken between 2015 and 2018 at wavelengths from 1.3 mm to 10 cm, probing pressures from $\sim$1 to $\sim$50 bar at spatial resolutions from 0.1'' to 0.8''.  Radiative transfer modeling was performed to determine the physical origin of the brightness variations across Uranus's disk. The radio-dark equator and midlatitudes of the planet (south of $\sim$50$^\circ$ N) are well fit by a deep H$_2$S mixing ratio of $8.7_{-1.5}^{+3.1}\times10^{-4}$ ($37_{-6}^{+13}\times$ Solar) and a deep NH$_3$ mixing ratio of  $1.7_{-0.4}^{+0.7}\times10^{-4}$ ($1.4_{-0.3}^{+0.5}\times$ Solar), in good agreement with literature models of Uranus's disk-averaged spectrum. The north polar region is very bright at all frequencies northward of $\sim$50$^\circ$N, which we attribute to strong depletions extending down to the NH$_4$SH layer in both NH$_3$ and H$_2$S relative to the equatorial region; the model is consistent with an NH$_3$ abundance of $4.7_{-1.8}^{+2.1} \times 10^{-7}$ and an H$_2$S abundance of $<$$1.9\times10^{-7}$ between $\sim$20 and $\sim$50 bar. Combining this observed depletion in condensible molecules with methane-sensitive near-infrared observations from the literature suggests large-scale downwelling in the north polar vortex region from $\sim$0.1 to $\sim$50 bar. The highest-resolution maps reveal zonal radio-dark and radio-bright bands at 20$^\circ$S, 0$^\circ$, and 20$^\circ$N, as well as zonal banding within the north polar region. The difference in brightness is a factor of $\sim$10 less pronounced in these bands than the difference between the north pole and equator, and additional observations are required to determine the temperature, composition and vertical extent of these features.

\end{abstract}


\section{Introduction}
\label{section_intro}

Uranus's 82$^\circ$ obliquity leads to drastic seasonal variations in insolation, with both poles receiving more annual sunlight than the equator. In addition, Uranus is the only giant planet that lacks an apparent internal heat source \citep{pearl90}. The unusual pattern of heat flux into the Uranian troposphere resulting from these two characteristics provides an extreme test of our understanding of atmospheric circulation \citep[for recent reviews see][]{hueso19, fletcher20}. The strong seasonal forcing also plays a role in altering Uranus's atmospheric composition; for example, Uranus's insolation pattern has been invoked to explain the disequilibrium in the H$_2$ ortho-para fraction in the upper troposphere as well as the seasonal variation in haze properties and/or methane abundance in the stratosphere \citep{hueso19}. Radio observations provide a unique tool for probing the atmosphere of Uranus beneath its tropospheric cloud layers, permitting inferences about its tropospheric properties \citep[][]{jaffe84, depater88, hofstadter89, hofstadter90, depater91, hofstadter92, hofstadter03, klein06}. 

Central to these questions is Uranus's global circulation pattern. Remote-sensing observations probe atmospheric vertical motions indirectly by determining the zonal-mean distribution of condensible gases and clouds. Regions of large-scale upwelling are cloudy and rich in condensible gases up to the condensation pressure of that gas, while downwelling regions are drier and less cloudy. This point can be understood by analogy to the Hadley cell on Earth \citep[see, e.g.,][]{marshall89}. Water vapor evaporated from the deep reservoir (the ocean) moves equatorward into the Intertropical Convergence Zone (ITCZ), where it upwells, condenses as the tropospheric temperature decreases with altitude, and then rains back into the deep reservoir. The air leaving the top of the ITCZ has thus been robbed of its water vapor by the tropopause cold trap, so the divergent upper branch and subsiding sub-tropical branch of the circulation cell are much drier than the ITCZ both in terms of relative humidity and column-integrated water vapor. The atmospheres of the giant planets organize into a series of alternating thermally direct and thermally indirect circulation cells similar to Earth's Hadley and Ferrel cells, giving rise to the spectacular jets of Jupiter and Saturn. Global circulation models (GCMs) of Jupiter show that the zonal-mean column-integrated abundances of both ammonia and water are indeed higher in the upwelling branches and lower in the downwelling branches \citep{young19a, young19b}, in agreement with ground-based \citep[e.g.,][]{depater16, depater19} and spacecraft \citep{li17} data. Based on Voyager thermal-infrared measurements, \citet{flasar87} suggested a circulation model for Uranus with gas rising at latitudes near 30$^\circ$ and subsiding at the equator and poles.  Observations with the Very Large Array (VLA) revealed a bright south pole on Uranus, interpreted as a relative lack of microwave opacity in the deep troposphere down to $\sim$50 bar, which pointed to large-scale subsidence of dry air \citep{jaffe84, depater88, depater89, hofstadter90, depater91, hofstadter92, hofstadter03, klein06}. A polar depletion in both methane and hydrogen sulfide (H$_2$S) at higher altitudes observed at visible and near-infrared wavelengths corroborated this interpretation \citep{karkoschka09, sromovsky14, irwin18, sromovsky19}. Together, these multi-wavelength observations lent further support to the \citet{flasar87} non-seasonal equator-to-pole single-cell meridional circulation pattern \citep[][]{depater91, hofstadter92, sromovsky14}. However, a single-celled model has difficulty explaining the observed bands of clouds at 38$^\circ$ and 58$^\circ$ in both hemispheres. An alternative model prescribing circulation in three vertically stacked layers was proposed by \citet[][]{sromovsky14}, although those authors make clear that the three-layer model has shortcomings of its own.  It also remains unclear how the bright storm systems observed occasionally at near-infrared wavelengths, which appear to migrate in latitude \citep{sromovsky07, depater11, depater15}, fit into either model.

Previously published radio and millimeter observations of Uranus \citep[e.g.][]{jaffe84, depater88}, mostly taken prior to 1990, have mostly considered the disk-integrated planet and primarily imaged Uranus's southern hemisphere. More recent spatially resolved observations using the VLA have been presented at conferences \citep[e.g.,][]{hofstadter09} and featured in a recent white paper \citep[][]{depater18}. Both the \citet{hofstadter09} image from 2005 and the \citet{depater18} image from 2015, which is analyzed in detail in this paper, show alternating bright and dark zonal bands in the midlatitudes, indicating a more complex circulation pattern than current models. The bright polar regions display similar brightness temperatures and zonal extents in the 2005 image, in which both poles are visible, and in the 2015 image zonal banding is visible within the bright polar region.

Inferences about the vertical cloud structure of giant planets are made by comparing radiative transfer and chemical modeling to observational data across the electromagnetic spectrum.
Visible and near-infrared spectroscopy \citep[e.g., ][]{karkoschka09, tice13, sromovsky14, dekleer15, sromovsky19} have identified methane as the major condensible species in the upper troposphere, producing bright clouds and weather readily observed in visible/IR imaging \citep[e.g., ][]{depater11, depater15, sromovsky15}. 
At $\sim$35 bar, gaseous NH$_3$ and H$_2$S are expected to precipitate into a cloud of solid NH$_4$SH, effectively removing either all nitrogen or all sulfur from the upper atmosphere. 
\citet{gulkis78} showed that Uranus must be ammonia-poor above the NH$_4$SH layer in order to fit the planet's disk-integrated radio spectrum. This finding suggested, contrary to solar composition models, that more H$_2$S than NH$_3$ was present in Uranus's deep atmosphere. Further work indicated that H$_2$S itself was also a major absorber, and new models, still with more sulfur than nitrogen, were developed that improved the fit to the observed spectra \citep[][]{depater91, depater18n}. Based on these studies, the uniform cloud layer at $\sim$3 bar, evident in IR spectroscopy, was long assumed to be composed of H$_2$S ice particles; this hypothesis was confirmed recently by the direct detection of H$_2$S spectral lines above the cloud \citep{irwin18}. 
The chemistry becomes more speculative at pressures deeper than $\sim$50 bars, as those depths have not been accessed observationally. Models suggest Uranus's oxygen is locked in a water-ice cloud at $\sim$270 K ($\sim$50 bar), and beneath this, an aqueous nitrogen-, sulfur-, and oxygen-bearing solution is expected to form \citep[][]{weidenschilling73, atreya85}.  The effects of phosphine (PH$_3$) have also been considered in the ice giants. Its condensation pressure is near 80 K ($\sim$1 bar on Uranus), and modeling shows that its absorption at deeper layers may be important to Uranus's infrared and millimeter spectrum \citep[e.g., ][]{fegley86, hoffman01}, especially near its $J = 1 \rightarrow 0$ rotational transition at 1.123 mm (266.9 GHz). However, its presence has never been confirmed observationally \citep[][]{orton89, moreno09}.

In this paper we present new VLA and ALMA observations of Uranus's atmosphere from 2015 to 2018, representing the highest sensitivity and spatial resolution measurements of the planet at wavelengths from 1.3 mm to 10 cm.
In Section \ref{section_obs} we outline the observational techniques and data processing procedures used to produce science images of Uranus. We present our results in Section \ref{section_results}, including seasonal brightness trends, spatial variations in brightness temperature, and inferred properties of Uranus's troposphere as determined from radiative transfer modeling. Finally, we put our results in a broader context and provide concluding remarks in Section \ref{section_conclusions}.

\section{Observations and Data Reduction}
\label{section_obs}

We obtained observations of Uranus with the Karl G. Jansky Very Large Array (VLA) from 0.9-10 cm (3.0-33 GHz) in August 2015 and the Atacama Large (sub-)Millimeter Array (ALMA) from 1.3-3.1 mm (98-233 GHz) between December 2017 and September 2018. A table of observations is provided in Table \ref{obstable}. The data-reduction procedures are outlined in the following two subsections.

\begin{table}[H]
	\scriptsize
	\begin{tabular}{|c|c|c|c|c|c|c|c|c|c|}
	 & Wavelength & Frequency &  & On-Source & Resolution & Resolution &  Flux \& Gain & Phase & Sub-Obs \\
	Array & (mm) & (GHz) & UT Date & Time (min) & (arcsec) & (km) & Calibrator & Calibrator & Latitude ($^{\circ}$) \\ 
	\hline
	VLA & 95  & 3.2 & 2015-08-29 & 88 & 0.79 & 11000 & 3C48 & J0121+422 & 33 \\
	VLA & 51   & 5.8 & 2015-08-29 & 89 & 0.45 & 6300 & 3C48 & J0121+422 & 33 \\
	VLA & 30.   & 9.9 & 2015-08-29/30 & 150 & 0.26 & 3600 & 3C48 & J0121+422 & 33 \\
	VLA & 20.   & 15 & 2015-08-29/30 & 148 & 0.17 & 2400 & 3C48 & J0121+422 & 33 \\
	VLA & 9.1  & 33 & 2015-08-29/30 & 160 & 0.085 & 1200 & 3C48 & J0121+422 & 33 \\
	ALMA & 3.1 & 98 & 2017-12-03 & 42 & 0.19 & 2700 & J0238+1636 & J0121+1149 & 38 \\
	ALMA & 3.1 & 98 & 2017-12-06 & 42 & 0.19 & 2700 & J0238+1636 & J0121+1149 & 38 \\
	ALMA & 2.1 & 144 & 2017-12-27 & 22 & 0.21 & 3000 & J0238+1636 & J0121+1149 & 38 \\
	ALMA & 1.3 & 233 & 2018-09-13 & 24 & 0.29 & 4000 & J0237+2848 & J0211+1051 & 45 \\
	\end{tabular}
	\caption{Table of observations. Note that the VLA observations at 0.9 cm, 2 cm, and 3 cm were taken in two parts over two consecutive days.\label{obstable}}
\end{table}


\subsection{ALMA Data}

\begin{figure}
	\includegraphics[width = 1.0\textwidth]{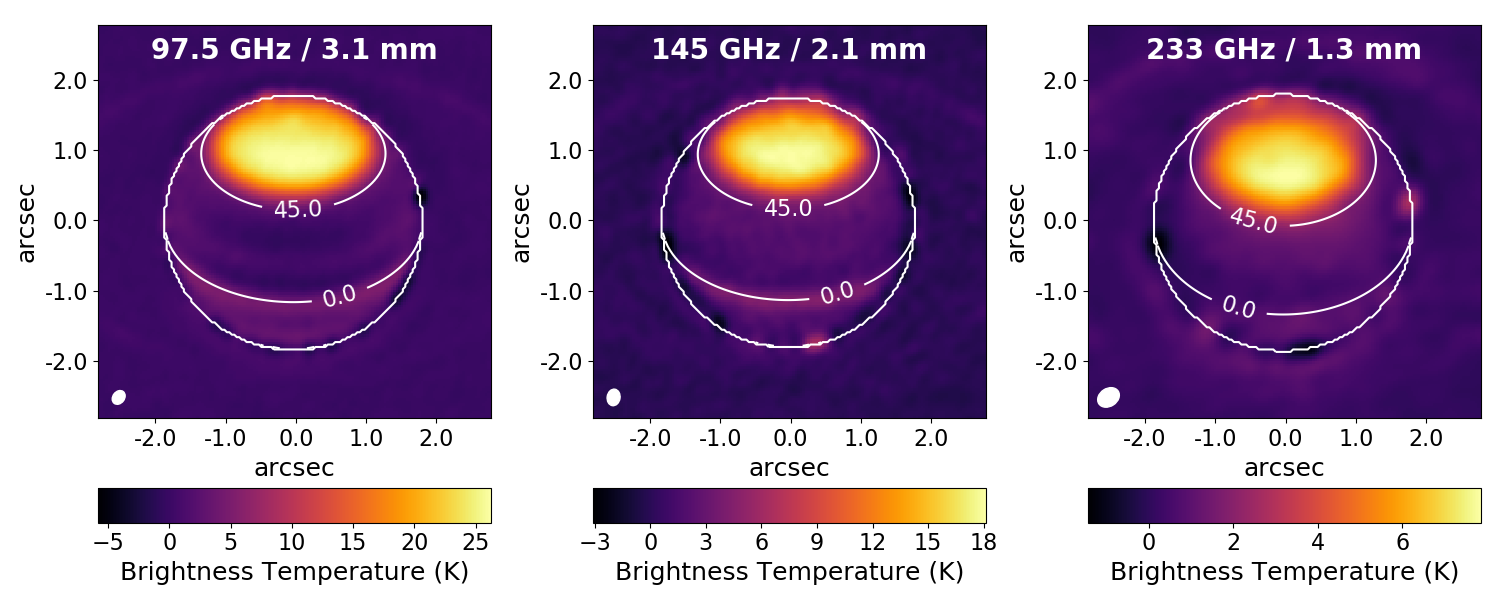}
	\includegraphics[width = 1.0\textwidth]{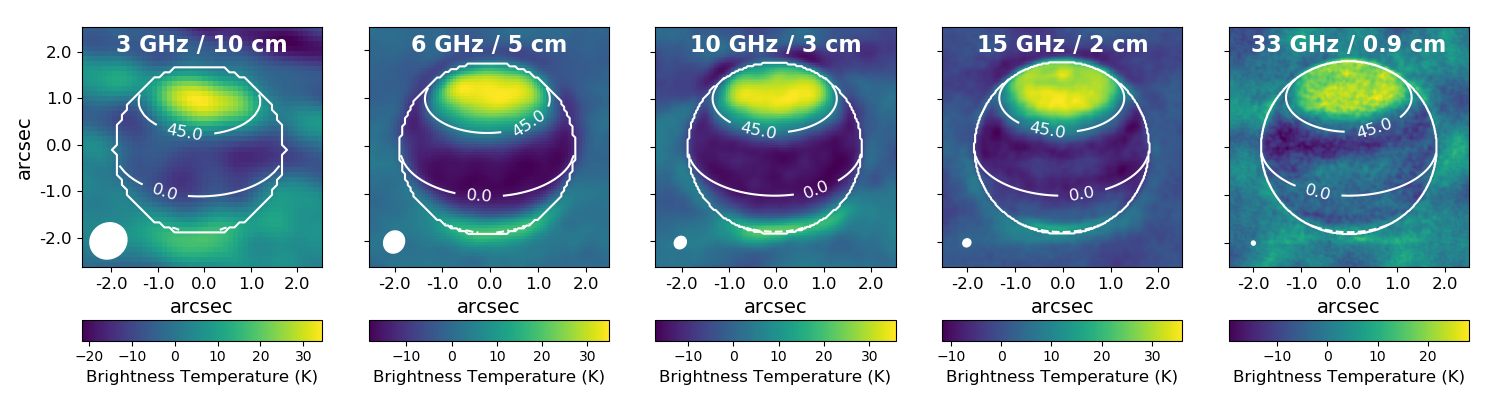}
	\caption{Disk-subtracted images of Uranus from ALMA \textbf{(top)} at 3.1 mm (Band 3), 2.1 mm (Band 4), and 1.3 mm (Band 6), and the VLA \textbf{(bottom)} from 0.9-20 cm. The color bars below each image indicate the brightness temperature residuals in Kelvin. The synthesized beam is shown as a white ellipse in the bottom left corner of each image. The ring of light and dark one-beam-size spots around the planet are artefacts produced by applying a Fourier transform to the UV-plane data near the sharp edges of the bright planet.\label{prettypics}}
\end{figure}

\begin{figure}
	\includegraphics[width = 0.5\textwidth]{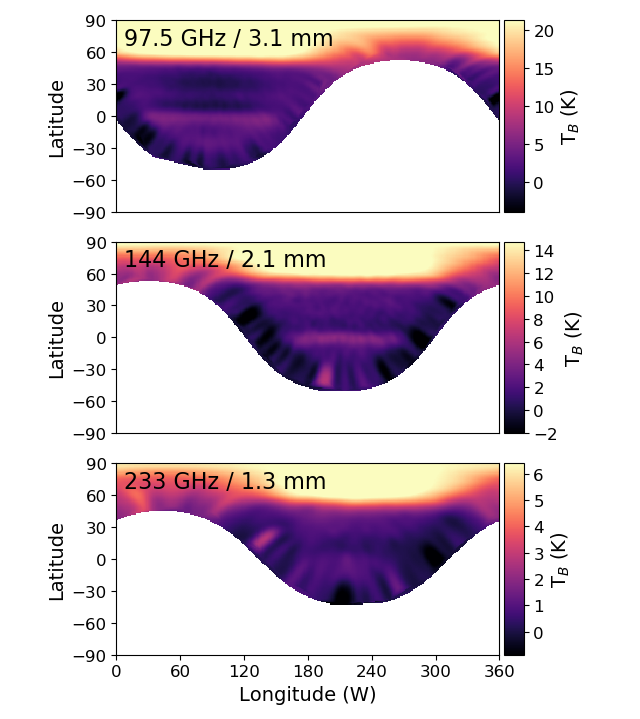}
	\includegraphics[width = 0.5\textwidth]{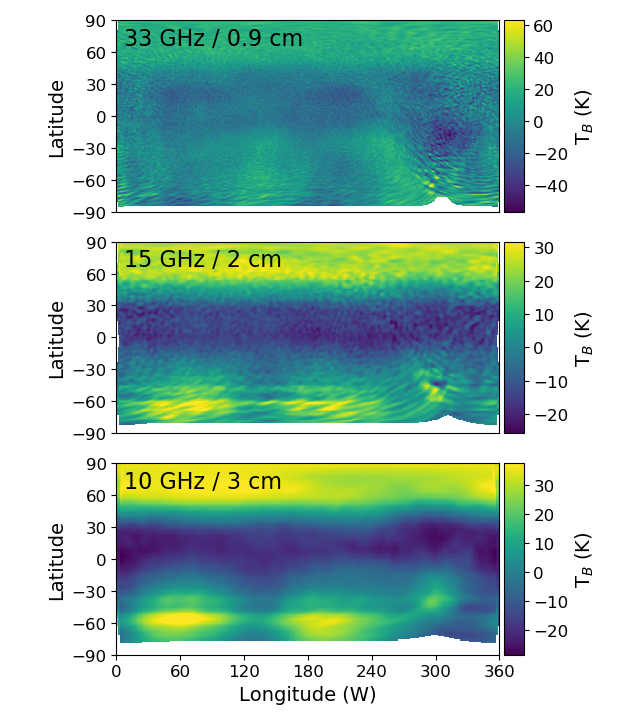}
	\caption{\textbf{Left:} Reprojected ALMA maps of Uranus. The $\sim$20 minute observations at 2.1 mm and 1.3 mm are smeared by $\sim$8$^\circ$, and the 3.1 mm  image is a sum of two $\sim$40 minute observations taken at different sub-observer longitudes. The alternating bright and dark spots near the planet's limb are due to CLEAN artefacts, induced by attempting to Fourier transform a sharp-edged planet and made larger in apparent size by projection effects. \textbf{Right:} Longitude-resolved VLA maps of Uranus, produced using the faceting technique \citep{sault04}. The distortions near 300$^\circ$W are due to poor zonal coverage at those longitudes in the observations. The spatial resolution of the 5 cm and 10 cm data is not sufficient to produce reliable longitude-resolved maps, so these are not shown.\label{projected}}
\end{figure}

The data in each of the three ALMA bands were flagged and calibrated by the North American ALMA Science Center using the standard data-reduction procedures contained in the NRAO's CASA software version 5.1.1. Standard flux- and phase-calibration procedures were carried out by applying the pipeline using the quasars listed in Table \ref{obstable} as calibrator sources. The CASA pipeline retrieved flux calibration errors of $\sim$5\% in all three bands. Iterative phase-only self-calibration, which is routinely applied to radio observations of bright planets \citep[e.g.,][]{butler01, depater14, depater19}, was performed using a procedure similar to that outlined in \citet{brogan18} using solution intervals of 20, 10, 5, and 1 minutes in that order. To reduce ringing in the image plane from the presence of the bright planet, a uniform limb-darkened disk model of Uranus was subtracted from the data in the UV plane in each band, as done in, e.g., \citet{depater14, depater16}. The disk-subtracted data were inverted into the image plane and deconvolved using CASA's \texttt{tclean} function.  The resulting disk-subtracted images are shown in Figure \ref{prettypics}. These images are also shown cylindrically projected onto a latitude-longitude grid in Figure \ref{projected}. It should be noted that the planet's rotation smears out features in longitude: the $\sim$20 minute observations at 2.1 mm and 1.3 mm are smeared by $\sim$8$^\circ$, and the 3.1 mm  image is a sum of two $\sim$40 minute observations taken at different sub-observer longitudes.

\begin{figure}
	\includegraphics[width = 0.7\textwidth]{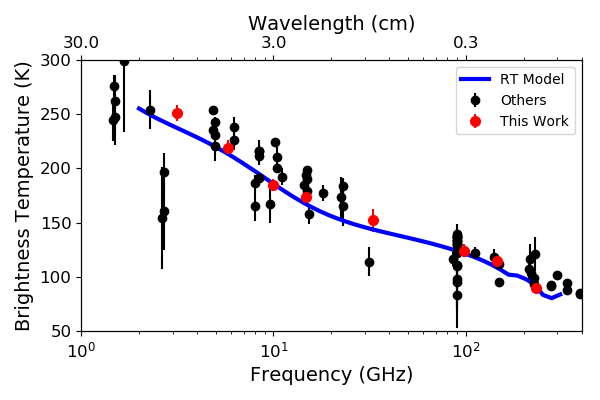}
	\caption{Disk-integrated brightness temperatures of Uranus from this work (red points) compared to measurements from the literature \citep[black points;][]{gulkis84, orton86, depater88, muhleman91, griffin93}. As shown by those authors (see also Figure \ref{kleinfig}), the scatter in the data is primarily due to real seasonal fluctuations in Uranus's observed brightness temperature as the planet's poles move into and out of view. Our data fall at a lower brightness temperature than the majority of the literature data because those were observed in southern summer, when the bright south polar region made up a large fraction of Uranus's disk. A radiative transfer model with parameters retrieved to match our 25$^\circ$N data (Section \ref{section_rt}), shown by a blue line, matches the disk-averaged data quite well.\label{planetFlux}}
\end{figure}

Images were also produced from the calibrated but non-disk-subtracted data; these images were used to measure absolute fluxes across the disk for radiative transfer modeling. The $\sim$3.5'' disk of Uranus was smaller than the maximum recoverable scale of the ALMA array configuration in all three observing bands\footnote{See the ALMA Technical Handbook for further discussion of the maximum recoverable scale https://almascience.nrao.edu/documents-and-tools/cycle5/alma-technical-handbook}; therefore we have short enough baselines to faithfully measure Uranus's total flux. These flux measurements were confirmed by fitting the visibility data to a Bessel function; the difference between the UV-plane-derived and image-plane-derived total flux measurements was much smaller than the flux calibration error at all wavelengths. The total flux measurements were corrected for the cosmic microwave background (CMB) according to the prescription detailed in Appendix A of \citet{depater14}. Final measurements of Uranus's disk-averaged brightness temperature are plotted against measurements from the literature in Figure \ref{planetFlux} and tabulated in Table \ref{tbs}. 

\begin{table}[H]
	\footnotesize
	\begin{tabular}{|c|c|c|c|c|c|c|c|}
	 Wavelength & Frequency & Disk-averaged & 25$^\circ$ N & 75$^\circ$ N & Flux Cal  & RMS per  & Latitude Bin \\
	(mm) & (GHz) & $T_B$ (K)     & $T_B$ (K)    & $T_B$ (K)    & Error (K) & Beam (K) & Error (K) \\ 
	\hline
	95 & 3.2 & 251.3 & 261.3 & 293.4 & 7.5  & 2.2 & 3.6 \\
	51  & 5.8 & 219.3 & 219.2 & 262.0 & 6.6  & 0.8 & 2.6 \\
	30.  & 9.9  & 184.5 & 182.8 & 224.4 & 5.5  & 1.2 & 2.6 \\
	20.  & 15  & 173.9 & 169.4 & 205.0 & 5.2  & 1.3 & 2.6 \\
	9.1   & 33  & 152.4 & 157.8 & 180.6 & 10.7 & 3.4 & 2.6 \\
	3.1 & 98  & 140.7 & 142.6 & 165.1 & 7.0  & 0.3 & 1.1 \\
	2.1 & 144  & 122.5 & 126.5 & 139.1 & 6.6  & 0.1 & 0.9 \\
	1.3 & 233 & 91.1  & 96.0  & 101.8 & 4.8  & 0.1 & 0.5 \\
	\end{tabular}
	\caption{Brightness temperature measurements and errors extracted from the ALMA and VLA maps. See Section \ref{section_obs} for discussion of the flux calibration and per-bin RMS errors, and Section \ref{spatial} for discussion of the latitude-bin error.\label{tbs}}
\end{table}

\subsection{VLA Data}

The data in each of the five VLA bands were flagged and calibrated using the standard data-reduction procedures contained in the MIRIAD software package\footnote{https://www.atnf.csiro.au/computing/software/miriad/} \citep[][]{sault95}. Standard flux- and phase-calibration procedures were carried out using the calibrator sources listed in Table \ref{obstable}; the absolute flux calibration error was assumed to be 3\% at the longest four wavelengths and 7\% at 0.9 cm \citep{butler01, perley13, perley17}. Iterative self-calibration, disk subtraction, imaging, and deconvolution were carried out in a similar manner to the ALMA data but using MIRIAD. The CMB correction was applied in the same way as for the ALMA images. The resulting disk-subtracted images are shown in Figure \ref{prettypics}. It should be noted that the 0.9 cm images are more strongly affected by imaging artefacts than the other bands because this wavelength sits near a telluric water absorption band, leading to relatively poor phase stability. Longitude-resolved images, produced using the faceting technique developed and recently used for VLA observations of Jupiter \citep{sault04, depater19}, are shown in Figure \ref{projected}. This represents the first attempt to resolve Uranus in longitude at these wavelengths. Limited UV-plane coverage leads to significant artefacts in these maps, including large-scale alternating bright and dark regions (e.g., bright at 60$^\circ$W and 200$^\circ$W in the 2 cm and 3 cm images) that cross many latitude bands. An apparent warping near 300$^\circ$W is also present, caused by poor zonal coverage at those longitudes. Nevertheless, if any vortex-like disturbances at the scale of Jupiter's Great Red Spot were present on Uranus, these maps should have detected them. No such structures are found, so in this work we focus our analysis on the longitude-smeared maps only.

\section{Results and Discussion}
\label{section_results}

\subsection{Seasonal Brightness Variations}
\begin{figure}
	\includegraphics[width = 0.52\textwidth]{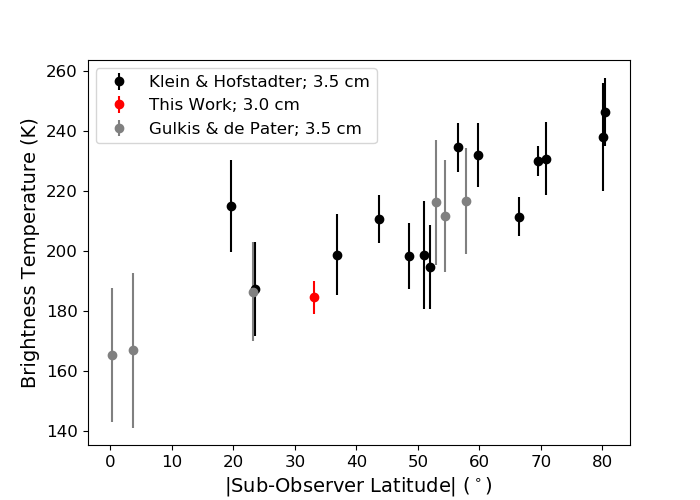}
	\includegraphics[width = 0.48\textwidth]{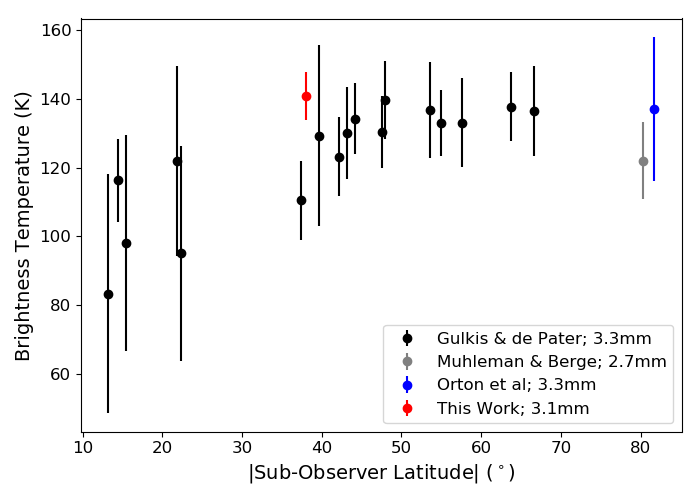}
	\caption{\textbf{Left:} VLA 3.0 cm total flux measurement of Uranus's disk from this work (red) compared to 3.5  cm flux measurements made with the Goldstone station of NASA’s Deep Space Network (DSN) at a range of sub-observer latitudes \citep{klein06}, as well as 3.5 cm measurements made prior to 1984 as summarized in \citet{gulkis84}.  \textbf{Right:} ALMA 3.1 mm total flux measurement of Uranus's disk from this work (red) compared to previous mm-wavelength measurements from various authors \citep{gulkis84, orton86, muhleman91}. The error bars on the literature measurements in both panels should be treated with caution, as flux calibration errors are often not reported. For the purposes of this plot we have added 5\% flux calibration errors to the points from the Gulkis and Muhleman papers. The x-axis in both panels shows the absolute value of sub-observer latitude; note that all previous measurements were made at negative sub-observer latitudes.\label{kleinfig}}
\end{figure}

We explore long-term trends in Uranus's radio brightness in Figure \ref{kleinfig}, which plots our VLA 3.0 cm data along with 3.5 cm data obtained by \citet{klein06} and older disk-averaged data from various telescopes \citep{gulkis84} as a function of sub-observer latitude. Our data were taken with Uranus's north pole facing the observer, whereas the \citet{klein06} data were taken toward the south pole; nevertheless, Uranus's brightness temperature was the same within the error bars of the data at a sub-observer longitude of $\sim$35$^\circ$. The $\sim$40 K brightness difference we observe between Uranus's northern midlatitudes and its north pole at 2-3 cm is of the same magnitude as VLA measurements of the difference between the south polar region and southern midlatitudes \citep[e.g., ][]{hofstadter90}. Imaging at 1.3 cm taken near equinox \citep[observers Hofstadter \& Butler; ][]{depater18}, which appears to show that both poles are roughly equally bright, further supports the similarity in brightness temperature between the north and south pole. Any trend in the 3 mm brightness temperature (right panel of Figure \ref{kleinfig}) is much less clear; the error bars are large for all observations taken near equinox, and all data taken at sub-observer latitudes larger than 40$^\circ$ have the same brightness to within the 2$\sigma$ level. This is confusing given the very large and bright north polar region we observe and the clear brightness temperature variations at longer wavelengths. However, the flux calibration uncertainty from measurements in some older papers \citep[see references in][]{gulkis84} have not been reported and so this result should be treated with caution.

\subsection{Spatially Resolved Brightness Temperatures}
\label{spatial}

\begin{figure}
	\includegraphics[width = 0.9\textwidth]{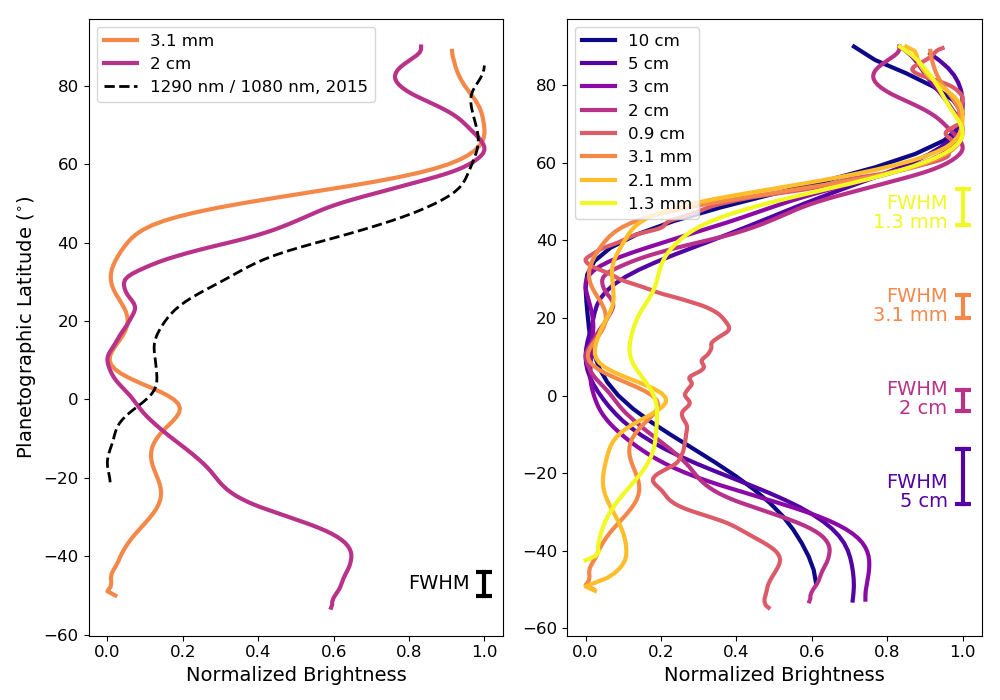}
	\caption{Meridional brightness profiles of the zonally-averaged radio/millimeter maps. \textbf{Left:} The 3.1 mm ALMA and 2.0 cm VLA data are plotted separately because they provide the best combination of high signal-to-noise ratio and resolution. These are compared to the brightness ratio between the Keck PaBeta (1.29 $\mu$m) and He1A (1.08 $\mu$m) filters, which is a tracer of the upper tropospheric methane abundance, as derived from 2015 imaging \citep[][; black dashed line]{sromovsky19}. The PaBeta/He1A ratio curve and VLA 2 cm curve have been convolved with a 1-D Gaussian beam at the ALMA 3.1 mm resolution (0.19 arcsec); the $\sim$6$^\circ$ FWHM of the beam at the sub-observer point is shown in the bottom right corner. \textbf{Right:} Latitudinal brightness profiles at all the observed wavelengths are plotted against one another. The FWHM of the beam at the sub-observer point is shown for representative frequencies in the same color as the data for that frequency. To facilitate visual comparison, the brightness temperature units in both panels are normalized so each band has maximum one.\label{latprofiles}}
\end{figure}

The images in Figures \ref{prettypics} and \ref{projected} reveal complex banding structure in Uranus's troposphere. For better visual comparison of the zonal features in the maps at different frequencies, a vertical slice of width 30$^\circ$ longitude centered on the sub-observer point was taken from the longitude-smeared, disk-subtracted, projected maps (shown in Figure \ref{projected} for the ALMA data) at each frequency, then averaged into a 1-D brightness versus latitude profile. The resulting profiles are shown normalized relative to one another in Figure \ref{latprofiles}. To extract spatially-resolved brightness temperatures for radiative transfer modeling, we simply averaged non-disk-subtracted latitude profiles over ten degrees latitude; that is, the 25$^\circ$N region represents latitudes from 20-30$^\circ$N and the 75$^\circ$N region represents latitudes from 70-80$^\circ$N. The error on the extracted brightness temperatures was determined by making latitude profiles over several different 30$^\circ$ longitude ranges and computing the standard deviation in the brightness measurements from those profiles. These ``latitude-bin'' errors are given in Table \ref{tbs}. It is worth noting that they are larger than the per-beam RMS error extracted from background regions of the radio maps (except at 0.9 cm). This is due to systematic errors arising from inverting and CLEANing visibility data of a very bright, extended, and sharp-edged source.


The north polar region is readily observed across the radio and millimeter spectrum as a prominent brightening northward of $\sim$50$^\circ$N. At 3.1 mm wavelength, this brightening reaches $>$25 K, nearly ten times the magnitude of any other brightness variations observed across Uranus's disk. The edge of this brightening appears sharply defined: the transition from the darker midlatitudes to the bright poles occurs over less than a single resolution element in all the observing bands. The magnitude of the polar brightening is much too large to be explained by variations in kinetic temperature (see Appendix \ref{appendix}), so we attribute it to downwelling air dry in absorbing species such as NH$_3$ and H$_2$S. Keck observations in the methane-sensitive PaBeta and H$_2$-sensitive He1A filters in 2015 \citep[][ simultaneous with our VLA observations]{sromovsky19} revealed a strong polar methane depletion that is spatially correlated with the bright polar region observed in the millimeter and radio data (see Figure \ref{latprofiles}). This implies continuously downwelling dry air over a wide range of pressures from 0.1 to at least 20 bar (see Section \ref{section_rt}) at latitudes northward of 50$^\circ$N. However, localized regions of upwelling (i.e. convection) or a temporally variable circulation pattern at the north pole cannot be ruled out. At 0.9 cm and 2 cm wavelengths, the highest-resolution VLA bands, a polar collar is observed at $\sim$60$^\circ$N, just north of the 45$^\circ$ polar collar seen at near-infrared wavelengths (see also Figure \ref{latprofiles}) but at the same latitude as the transition to solid-body rotation in Uranus's zonal wind profile \citep[][]{sromovsky12}. The polar collar is bounded by a somewhat fainter band at $\sim$75$^\circ$N, then another brightening right at the north pole. However, this banding is not observed in the ALMA data despite similar spatial resolutions in the 3 mm and 2 cm data.

Alternating bright and dark bands are observed near Uranus's equator in some of our images, with brighter latitudes at $\sim$20$^\circ$S, $\sim$0$^\circ$, and $\sim$20$^\circ$N. At 3.1 mm, these bands are $\sim$1 K, $\sim$4 K, and $\sim$1 K brighter, respectively, than their surroundings. The banding is visible in all the ALMA data (3.1 mm, 2.1 mm, and 1.3 mm). It is also observed faintly in the 2 cm and 0.9 cm VLA bands, but not at longer wavelengths. Using our radiative transfer model (see Section \ref{section_rt}), we find that variations in the CH$_4$ abundance, the PH$_3$ abundance, the relative humidity of H$_2$S, or the ortho-para fraction can all produce stronger absorption from 1 to 3 mm than at wavelengths longer than 2 cm. Changes in the kinetic temperature may also be responsible for all or part of the brightness temperature banding; however, this would require unusual atmospheric temperature profiles to fit our radio-millimeter spectrum (see Appendix \ref{appendix} for a discussion of the kinetic temperature as it relates to the polar region). These putative abundance variations suggest downwelling and depletion in condensing species at the brighter latitudes extending at least from $\sim$1 to 5 bar. The lack of strong banding at longer wavelengths means that either the depletions do not extend deeper than 5-10 bar, the depletions are caused by a species that absorbs more strongly at shorter wavelengths, or the longer-wavelength VLA observations lack the sensitivity and spatial resolution to detect these faint equatorial bands. These observations point to a more complex circulation pattern than predicted by models \citep[][]{flasar87, allison91, sromovsky14}, which suggested upwelling near 30$^\circ$N and 30$^\circ$S and subsidence at the equator and poles. A recent review paper \citep{fletcher20} considered a more complex model that prescribes tropospheric upwelling at the equator and just equatorward of the polar region ($\sim$40$^\circ$N), and downwelling between these. This model captures the large region of subsidence that we require at the poles. However, the alternating bright and dark bands we see, if indeed tied to upwelling and downwelling, do not match very well the locations prescribed in their model.

\subsection{Radiative Transfer Modeling}
\label{section_rt}

\begin{figure}
	\includegraphics[width = 0.5\textwidth]{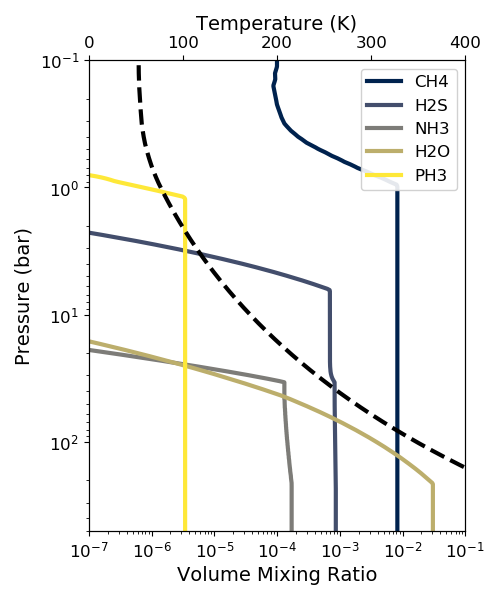}
	\caption{Vertical abundance profiles for trace gases in our radiative transfer model. Deep atmospheric abundances are set to 35$\times$ Solar except for NH$_3$, which has an abundance of 1$\times$ Solar. The dashed black line plots the assumed temperature-pressure profile, which follows a moist adiabat.\label{nominalmodel}}
\end{figure}

The combined ALMA and VLA datasets comprise a spectrum of Uranus. Here we use radiative transfer models to fit this spectrum and infer the vertical distribution of condensible species in Uranus's atmosphere. Radiative transfer modeling was carried out using the radio-BEAR (radio-BErkeley Atmospheric Radiative transfer) code\footnote{https://github.com/david-deboer/radiobear}. The cloud model and radiative transfer scheme are described in detail in \citet{depater05, depater14, depater19}. Radio-BEAR assumes that the atmosphere is in local thermodynamic equilibrium; its temperature follows an adiabat with a temperature of 76.4 K at 1 bar as determined by radio-occultation experiments with Voyager 2 \citep{lindal87}. At higher altitudes, where radiative effects become important, we used the temperature-pressure profile derived from Voyager/IRIS observations by \citet{orton15}. The temperature-pressure profile is shown in Figure \ref{nominalmodel}. Cloud densities may affect the millimeter and/or radio spectrum via absorption and scattering; however, too little is known about the cloud properties on Uranus to make an accurate cloud density model and clouds have been shown not to significantly affect the opacity at these wavelengths on Jupiter \citep{depater19}. RadioBEAR therefore ignores cloud opacity and considers only gas absorption.

\begin{figure}
	\includegraphics[width = 1.0\textwidth]{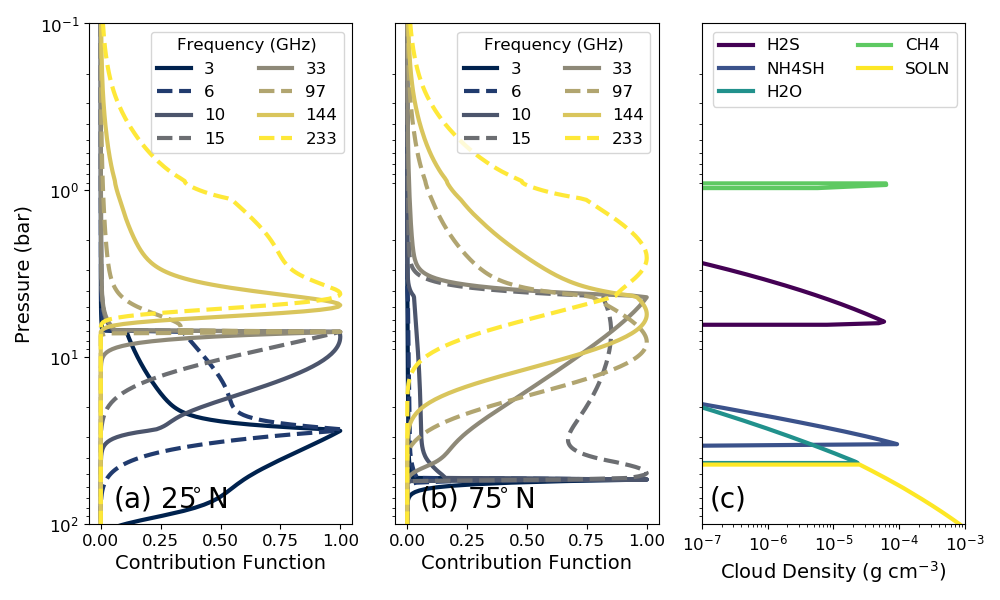}
	\caption{Normalized contribution function at each observed frequency for the best-fitting radiative transfer models at \textbf{(a)} 25$^\circ$N and \textbf{(b)} 75$^\circ$N. The spike in panel (b) at $\sim$40 bar is caused by the discontinuity in the vertical abundance profiles at the retrieved mixing pressure. \textbf{(c)} Cloud density as a function of pressure for clouds expected to form under thermochemical equilibrium assuming the abundance profiles in Figure \ref{nominalmodel}.\label{weighting}}
\end{figure}

We started with a nominal disk-averaged model that assumed deep H$_2$S, H$_2$O, and CH$_4$ abundances of 30$\times$ Solar and a deep NH$_3$ abundance of 1$\times$ Solar \citep{depater91, depater18}.\footnote{Solar abundances are assumed to be the protosolar values given in \citet{asplund09} C/H$_2$ = $5.90\times10^{-4}$; N/H$_2$ = $1.48\times10^{-4}$; O/H$_2$ = $1.07\times10^{-3}$; S/H$_2$ = $2.89\times10^{-5}$; Ar/H$_2$ = $5.51\times10^{-6}$; P/H$_2$ = $5.64\times10^{-7}$} The vertical profiles of these gases were determined using a cloud-physics model \citep[][]{atreya85, romani86, depater91} that includes prescriptions for a water-solution cloud at $\gtrsim$100 bar and an NH$_4$SH layer at $\sim$35 bar. We added in PH$_3$ with a nominal deep abundance of 30$\times$ Solar and a simple prescription for its saturation vapor curve \citep[][]{orton89}.
Vertical profiles of all the condensible gases included in the radiative transfer code are plotted in Figure \ref{nominalmodel}. The observed frequencies are sensitive to the atmospheric abundance and temperature from $\sim$0.1-50 bar; contribution functions at each frequency for the nominal model are shown in Figure \ref{weighting}.

\begin{figure}
	\includegraphics[width = 0.9\textwidth]{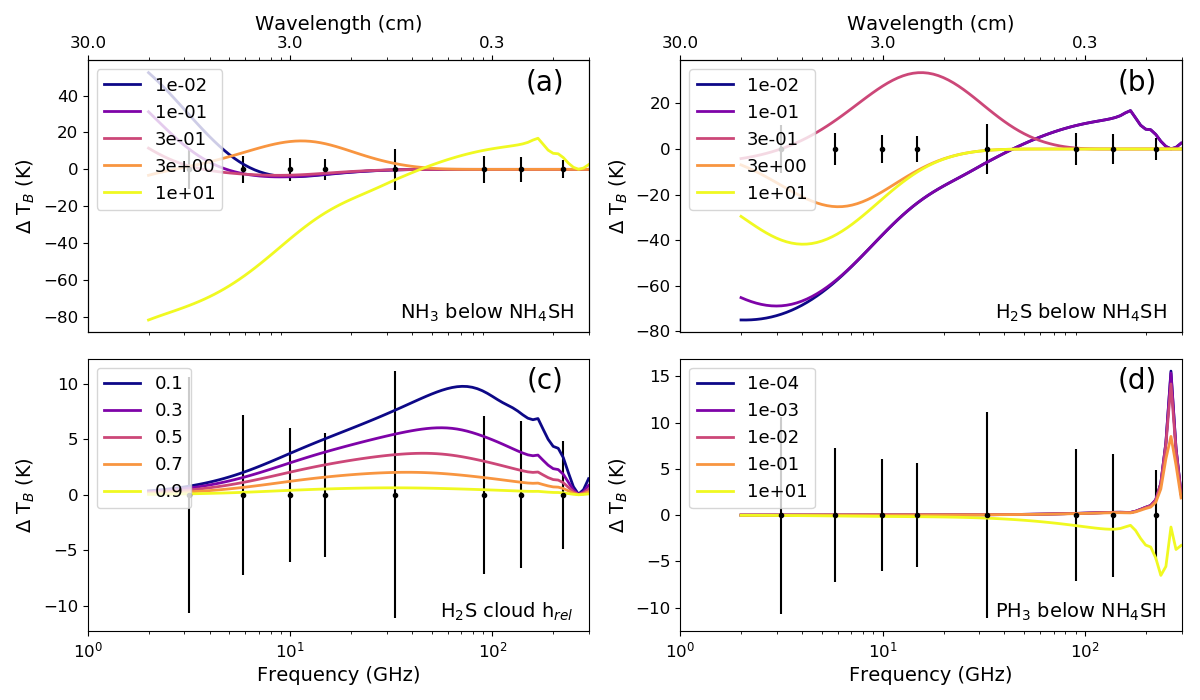}
	\caption{Effect of changing radiative transfer model parameters on the radio spectrum of Uranus. In each panel, one parameter was changed from its nominal value; the resulting model spectrum was subtracted from the nominal model such that $\Delta T_B$ represents the departure from the nominal model. Legend labels denote a multiplicative factor applied to the model parameter of interest, where 1.0 is the nominal value (30$\times$ Solar for H$_2$S and PH$_3$; 1$\times$ Solar for NH$_3$, 1.0 for H$_2$S h$_{rel}$). The black points plotted along the zero line show the size of the error bars on the data at the observed frequencies. All models were produced assuming viewing geometry from the sub-observer point.\label{varyparams}}
\end{figure}

We next perturbed many possible variables within this model one at a time to observe their effect on the spectrum, namely: the abundances of NH$_3$, H$_2$S, CH$_4$, and PH$_3$ from 500 bar up to just below the NH$_4$SH layer\footnote{This effectively ignores the solution cloud at $\sim$100 bar. However, we are mostly insensitive to the solution cloud's effect on the spectrum, so for practical purposes this is the same as varying the abundances just above the solution cloud. The reason we made this choice and its effects are discussed further in Sections \ref{section_enriched} and \ref{section_discussion}.}; the relative humidity of the H$_2$S ice cloud (H$_2$S $h_{rel}$); the ortho-para hydrogen fraction; and a wet vs dry adiabat. Only NH$_3$ and H$_2$S gas absorb strongly enough to impact the radio spectrum by $\gtrsim$5 K, i.e. above the flux calibration error, at the observed frequencies. We therefore expect strong constraints on only the H$_2$S and NH$_3$ abundances and the H$_2$S relative humidity; the impacts of these three parameters on Uranus's radio spectrum spectrum are shown in Figure \ref{varyparams}. The effect of the NH$_4$SH cloud can be seen on the spectrum as NH$_3$ and H$_2$S are varied (Figure \ref{varyparams}a and \ref{varyparams}b): at low H$_2$S abundances and high NH$_3$ abundances, NH$_3$ survives above the NH$_4$SH cloud and absorbs strongly at long wavelengths, whereas at high H$_2$S abundances and low NH$_3$ abundances (including the nominal values), H$_2$S survives above the NH$_4$SH cloud and absorbs most strongly at millimeter wavelengths. In the intermediate regime, where the NH$_3$ and H$_2$S abundances are nearly equal, the lower H$_2$S abundance leads to less absorption from 1-3 cm, but at 10 cm the NH$_3$ absorption deeper than the NH$_4$SH cloud is still important (see salmon line in Figure \ref{varyparams}b). Pressure-broadened absorption near the 1.123 mm (266.9 GHz) rotational line of phosphine may also be important in ALMA Band 6, i.e. at 1.3 mm (see Figure \ref{varyparams}d). Methane has no strong lines at millimeter or radio wavelengths, but alters Uranus's millimeter/radio spectrum at the $\sim$1$\sigma$ level primarily by altering the strength of H$_2$-CH$_4$ collision-induced absorption \citep[CIA;][]{borysow86}.  The ortho-para fraction of molecular hydrogen changes the strength of H$_2$-H$_2$ CIA as well as modifying the adiabatic lapse rate, and thus also affects the mm/radio spectrum \citep[][]{trafton67, wallace80}.\footnote{The default ortho-para hydrogen state is ``normal'' H$_2$, which denotes high-temperature-limit value of 3:1 orthohydrogen:parahydrogen, which is reached near 300 K.  ``Equilibrium" H$_2$ refers to the equilibrium ortho-para fraction at the temperature of each atmospheric layer according to the T-P profile in Figure \ref{nominalmodel}. The ortho-para fraction may achieve disequilibrium due to vertical mixing, since the timescale to convert between the ortho- and para-states is much longer than dynamical timescales \citep[][]{trafton67, wallace80}.} Since in our model the CH$_4$ abundance, PH$_3$ abundance, and ortho-para fraction of H$_2$ may all affect the spectrum to near the $\sim$1$\sigma$ level at some frequencies, we allow them to vary as well. The deep H$_2$O abundance has no impact on the spectrum because the H$_2$O cloud forms well below the maximum depth to which we are sensitive; we set the deep H$_2$O abundance to a fixed 30$\times$ Solar. The difference between a wet and dry adiabat is small; we assumed a dry adiabat.

Radiative transfer modeling was carried out within a Markov Chain Monte Carlo (MCMC) framework, implemented using the \texttt{emcee} Python package \citep[][]{dfm13}.\footnote{https://emcee.readthedocs.io/en/v2.2.1/}
Letting $\theta$ represent the set of free parameters in the model, the likelihood function $\ln p$ is given by
\begin{equation}
	\label{lnlike}
	\ln p(T|\nu, \sigma, \theta) = -\frac{1}{2} \sum_n \Big[ (T_n - T_m(\theta))^2 \sigma_n^{-2} + \ln(2 \pi \sigma_n^{-2}) \Big]
\end{equation}
where $\sigma_n^2$ is the variance of the measured brightness temperature $T_n$ at each frequency $\nu_n$. At each MCMC step, a set of test parameters $\theta$ is selected, a model brightness temperature $T_m(\theta)$ is generated by RadioBEAR at each frequency, and the likelihood function is evaluated. The result of many MCMC iterations is a joint probability distribution over the free parameters in the radiative transfer model. Each of the MCMC runs presented in this paper used 500 iterations and 40 walkers. As is standard practice with MCMC \citep{dfm13}, we cut out the ``burn-in'' phase by using only the second half of the iterations to describe the posterior distribution; plotting the parameter values as a function of iteration confirmed that this procedure had worked as intended. We refer the interested reader to \citet{hogg18} for an overview of how MCMC works and is used in astrophysics research.

\subsubsection{Enriched Region}
\label{section_enriched}

\begin{figure}
	\includegraphics[width = 1.0\textwidth]{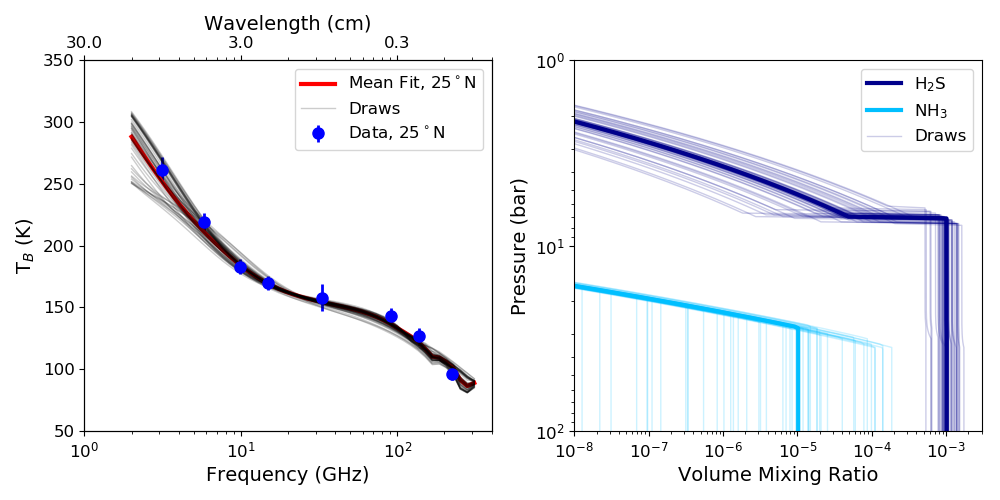}
	\includegraphics[width = 1.0\textwidth]{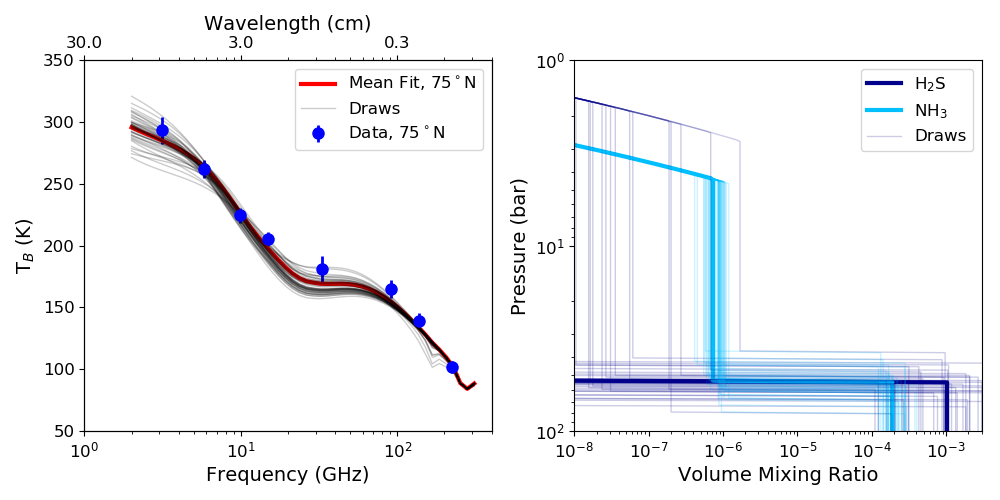}
	\caption{\textbf{Left:} Brightness temperature measurements of Uranus (blue dots) compared to our radiative transfer models at 25$^\circ$N (top) and 75$^\circ$N (bottom). The best-fitting model is shown as a thick red line, and 50 MCMC draws are shown as thin grey lines. \textbf{Right:} Abundance profiles of H$_2$S (dark blue) and NH$_3$ (light blue) for the best-fitting model (thick line) and the same 50 draws (thin lines) at 25$^\circ$N (top) and 75$^\circ$N (bottom). We refer the reader to \citet{hogg18} for an explanation of the meaning of model draws.\label{bestfit}}
\end{figure}

\begin{figure}
	\includegraphics[width = 0.9\textwidth]{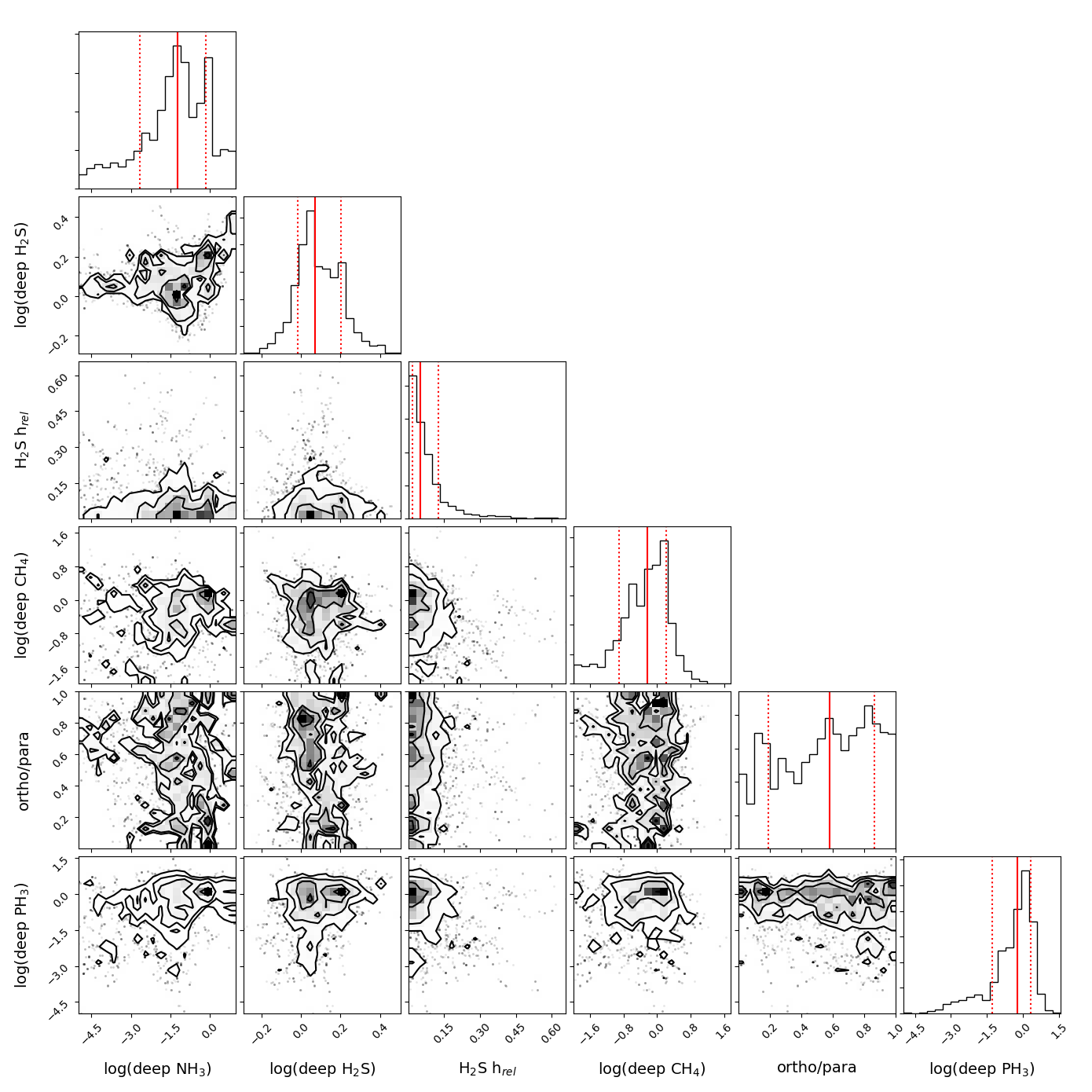}
	\caption{``Corner plot'' showing the one-dimensional (top panels, corresponding to the label at the bottom of each column) and two-dimensional (other panels) projections of the posterior probability
distribution of the MCMC-retrieved parameters for the radiative transfer models at 25$^\circ$N. The mean value (solid red line) and 16th and 84th percentile (dotted red lines) of each probability distribution are plotted. We refer the reader to \citet{hogg18} for an explanation of how to interpret a corner plot.\label{corners1}}
\end{figure}

We first fit the radio-dark region at 25$^\circ$N, chosen as the representative ``enriched'' or volatile-rich region because of its location at small emission angles with respect to the observer as well as relatively similar brightnesses at nearby latitudes, minimizing beam smearing effects. For computational efficiency, we simplified the full cloud-physics model to include only prescriptions for the NH$_4$SH cloud and the H$_2$S, NH$_3$, CH$_4$, and PH$_3$ ice clouds. The solution cloud beneath the NH$_4$SH cloud was ignored; this was a reasonable compromise because its location at $\sim$100 bar pressures is deeper than the observed frequencies probe (see Figure \ref{weighting}). The vertical abundance profile of H$_2$O as well as all other inputs (ortho-para fraction, adiabat) were set to their nominal values.  We checked our simplified implementation against the full cloud-physics model, and found that the difference in brightness temperature between them was at least an order of magnitude smaller than the flux calibration errors at all frequencies. However, it should be noted that the MCMC retrieval estimates the NH$_3$ and H$_2$S abundances below the NH$_4$SH layer but above the solution cloud, i.e. at $\sim$50-100 bar. The solution cloud in the full model removes $\sim$5\% of the deep H$_2$S and $\sim$25\% of the deep NH$_3$; this is discussed further in Section \ref{section_discussion}. The retrieved values for all the parameters at 25$^\circ$N are given in Table \ref{vals25}. The best-fitting model is compared to the data in Figure \ref{bestfit}, and a ``corner plot'' displaying the one- and two-dimensional projections of the posterior probability distribution of the retrieved parameters is shown in Figure \ref{corners1}. The observed disk-averaged brightness temperatures are also reasonably well fit by the 25$^\circ$ model (see Figure \ref{planetFlux}).


\begin{table}[H]
	\footnotesize
	\begin{tabular}{|c||c|c|c|c|c||c|}
	

	Parameter & 2.5\% ($-2\sigma$) & 16\% ($-1\sigma$) & Median & 84\% ($+1\sigma$) & 97.5\% ($+2\sigma$) & Abundance / Solar (1$\sigma$) \\
	\hline
	NH$_3$ below NH$_4$SH & $4.07\times10^{-9}$ & $2.63\times10^{-7}$ & $7.23\times10^{-6}$ & $8.49\times10^{-5}$ & $7.40\times10^{-4}$ & 0.06$^{+0.6}_{-0.04}$ \\
	H$_2$S below NH$_4$SH & $5.34\times10^{-4}$ & $6.88\times10^{-4}$ & $8.27\times10^{-4}$ & $1.12\times10^{-3}$ & $1.50\times10^{-3}$ & 35.2$^{+12.5}_{-5.9}$ \\
	H$_2$S h$_{rel}$ & 0.005 & 0.02 & 0.05 & 0.13 & 0.34 & ---  \\
	CH$_4$ below NH$_4$SH & $2.68\times10^{-4}$ & $1.77\times10^{-3}$ & $8.29\times10^{-3}$ & $2.39\times10^{-2}$ & $5.73\times10^{-2}$ & 17$^{+33}_{-14}$ \\
	ortho-para & 0.03 & 0.19 & 0.58 & 0.87 & 0.97 & --- \\
	PH$_3$ below NH$_4$SH & $4.52\times10^{-9}$ & $2.86\times10^{-7}$ & $3.43\times10^{-6}$ & $1.16\times10^{-5}$ & $2.72\times10^{-5}$ & 18$^{+43}_{-17}$ \\
	\hline
	H$_2$S above NH$_4$SH & $4.29\times10^{-4}$ & $6.32\times10^{-4}$ & $8.09\times10^{-4}$ & $1.00\times10^{-3}$ & $1.31\times10^{-3}$ & 34.4$^{+8.3}_{-7.7}$ \\
	Deep H$_2$S & $5.59 \times 10^{-4}$ & $7.20 \times 10^{-4}$ & $8.65 \times 10^{-4}$ & $1.17 \times 10^{-3}$ & $1.57 \times 10^{-3}$ & 36.8$^{+13.1}_{-6.2}$ \\
	\end{tabular}
	\caption{Values and errors for radiative transfer model parameters for the ``enriched'' region at 25$^\circ$N. The median values are reported, along with the 16th/84th percentile values, which represent the 1$\sigma$ interval, and the 2.5/97.5 percentile values, which represent the 2$\sigma$ interval. The quantities below the horizontal line are not free parameters in the model, but can be determined from the retrieved vertical profiles. The values below the solution cloud are measured at 35 bar. The ortho-para fraction takes a value between 0 and 1, where 1 is ``equilibrium'' H$_2$ and 0 is ``normal'' H$_2$. The deep H$_2$S abundance refers to the model abundance below the water solution cloud.\label{vals25}}

\end{table}

\begin{figure}
	\includegraphics[width = 0.9\textwidth]{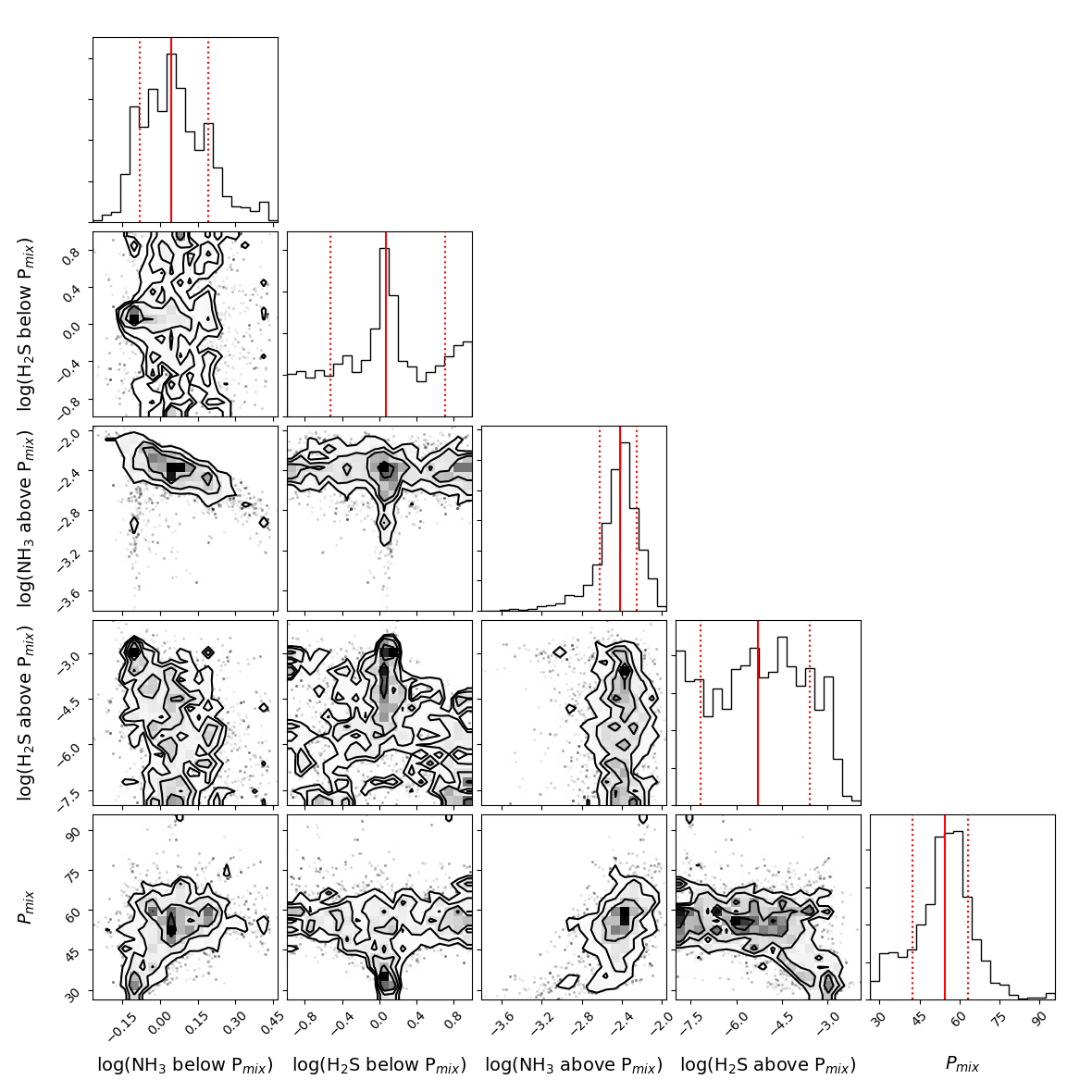}
	\caption{Same as Figure \ref{corners1} but for the depleted model and data at 75$^\circ$N.\label{corners2}}
\end{figure}

\begin{figure}
	\includegraphics[width = 0.7\textwidth]{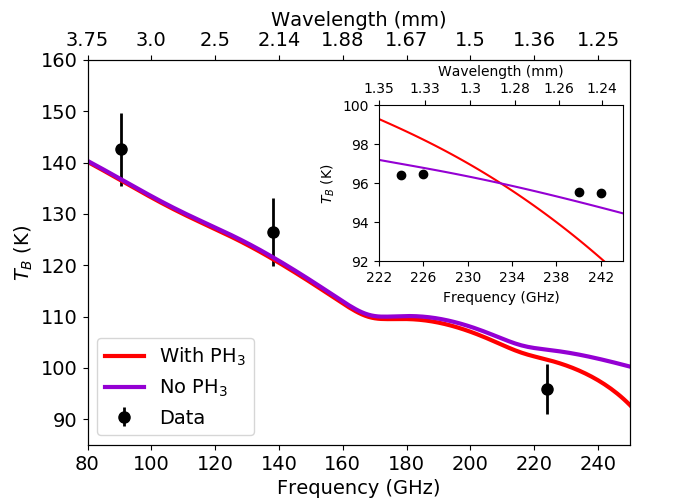}
	\caption{\textbf{Main:} ALMA data at 25$^\circ$N (black points) plotted over radiative transfer models with 35$\times$ Solar phosphine (red line) and without any phosphine (purple line). The inclusion of phosphine marginally improves the fit. In this plot the error bars include the flux calibration error and are identical to those in the left panel of Figure \ref{bestfit}. \textbf{Inset:} ALMA data in Band 6 (1.24-1.35 mm) split into its four spectral windows. We plot here only the error due to random noise, which is very small compared to the brightess of Uranus; the error bars are of the same order as the thickness of the data points. The radiative transfer models are normalized so that their value at 1.29 mm (233 GHz) is equal to the mean of the Band 6 data, to facilitate visual comparison of the spectral slopes. Here the inclusion of phosphine makes the spectral slope agreement poorer.\label{band6_spectralslope}}
\end{figure}

The abundance of ammonia below the NH$_4$SH cloud is only weakly constrained in the enriched region, with a best-fitting value of 6\% Solar but a 1$\sigma$ confidence interval from 0.2\% to 70\% Solar. This quantity is better constrained in the depleted north polar region, where the very low H$_2$S absorption allows our observations to probe deeper into the atmosphere (see Section \ref{npspot}).
The hydrogen sulfide abundance takes a 1$\sigma$ value of $35.2_{-5.9}^{+12.3} \times$ Solar below the NH$_4$SH cloud, in excellent agreement with previous results \citep[][]{depater91}. The relative humidity of the H$_2$S ice cloud takes an 84th percentile value of 0.13, meaning that $h_{rel} < 13$\% is preferred to higher values.
Models that include at least some PH$_3$ are preferred, as absorption from the pressure-broadened PH$_3$ line at 1.123 mm (266.9 GHz) is the only way in our model to decrease the brightness temperature at 1.3 mm relative to 2.1 mm. Indeed, a PH$_3$ abundance below $\sim$3\% Solar is disfavored at the 2$\sigma$ level. However, the spectral slope within the 1.3 mm ALMA band is more consistent with models lacking phosphine absorption than models including it (see Figure \ref{band6_spectralslope}). A detailed study of Uranus's radio spectrum around the PH$_3$ $J = 1 \rightarrow 0$ absorption line at 1.123 mm (266.9 GHz) is required to confirm or rule out the presence of significant PH$_3$ in the Uranian troposphere.
The methane abundance is only weakly constrained. Our retrieved abundance agrees with previous measurements at near-infrared and visible wavelengths, which range from $\sim$3 to $\sim$5\% in the equatorial regions \citep{karkoschka09, tice13, sromovsky14, sromovsky19, irwin19}, but is not as constraining as these studies.
The average spin state of hydrogen (ortho-para) remains completely unconstrained by our data.

\subsubsection{North Pole}
\label{npspot}

We then modeled the brightness temperature of the depleted north polar region at 75$^\circ$ N. Physically, the situation in the downwelling depleted region differs from that in the upwelling enriched region: the subsiding air is already depleted in condensible species, so strong subsaturations are likely at certain pressures. In addition, the depletion in abundances does not extend to infinite depth---the atmosphere must be well-mixed deeper than some pressure P$_{mix}$. Thus the deep abundance should remain the same as retrieved in the enriched region. With these considerations in mind, we employ a model for the depleted region that prescribes a step function in the H$_2$S and NH$_3$ abundances. We set the prior probability distribution functions of the deep abundances equal to the posterior probability distribution functions found for the enriched region; that is, the deep abundances are constrained to agree with the enriched region.  The H$_2$S and NH$_3$ abundances are uniform above some mixing pressure $P_{mix}$ until the species reaches its condensation level; the depletion factor of each species and $P_{mix}$ are allowed to vary freely. The retrieved values for all the parameters at 75$^\circ$N are given in Table \ref{vals75}. The best-fitting model is compared to the data in Figure \ref{bestfit}, and a ``corner plot'' displaying the one- and two-dimensional projections of the posterior probability distribution of the retrieved parameters is shown in Figure \ref{corners2}.

\begin{table}[H]
	\footnotesize
	\begin{tabular}{|c||c|c|c|c|c||c|}
	
	
	Parameter & 2.5\% ($-2\sigma$) & 16\% ($-1\sigma$) & Median & 84\% ($+1\sigma$) & 97.5\% ($+2\sigma$) & Abundance / Solar (1$\sigma$) \\
	\hline
	NH$_3$ below P$_{mix}$ & $8.46\times10^{-5}$ & $1.00\times10^{-4}$ & $1.31\times10^{-4}$ & $1.86\times10^{-4}$ & $2.74\times10^{-4}$ & 1.10$^{+0.46}_{-0.26}$ \\
	H$_2$S below P$_{mix}$ & $8.46\times10^{-5}$ & $2.07\times10^{-4}$ & $8.46\times10^{-4}$ & $3.61\times10^{-3}$ & $6.41\times10^{-3}$ & 36$^{+118}_{-27}$ \\
	NH$_3$ above P$_{mix}$ & $1.09\times10^{-7}$ & $2.88\times10^{-7}$ & $4.67\times10^{-7}$ & $6.75\times10^{-7}$ & $9.53\times10^{-7}$ & $3.9^{+1.7}_{-1.5}\times10^{-3}$ \\
	H$_2$S above P$_{mix}$ & --- & --- & --- & $1.89\times10^{-7}$ & $1.11\times10^{-6}$ & $<$0.008 \\
	$P_{mix}$ & 30.64 & 42.23 & 54.59 & 63.23 & 75.58 & --- \\
	\hline
	Deep NH$_3$ & $1.02\times10^{-4}$ & $1.31\times10^{-4}$ & $1.71 \times 10^{-4}$ & $2.43 \times 10^{-4}$ & $3.92 \times 10^{-4}$ & 1.44$^{+0.60}_{-0.34}$ \\
	\end{tabular}
	\caption{Values and errors for radiative transfer model parameters for the ``depleted'' north polar region at 75$^\circ$N. The median values are reported, along with the 16th/84th percentile values, which represent the 1$\sigma$ interval, and the 2.5/97.5 percentile values, which represent the 2$\sigma$ interval. The quantities below the horizontal line are not free parameters in the model, but can be determined from the retrieved vertical profiles. The deep NH$_3$ abundance refers to the model abundance below the H$_2$O solution cloud.\label{vals75}}
\end{table}

The MCMC simulation shows that strong depletions in both NH$_3$ and H$_2$S are required above the mixing pressure.
The results are consistent with zero H$_2$S above the mixing pressure; the simulation yields a 1$\sigma$ upper limit of 0.8\% Solar, which translates to at least $\sim$4000 times less H$_2$S above $P_{mix}$ than in the enriched region, i.e., an abundance $\lesssim$$2\times10^{-7}$. The results do not constrain the H$_2$S abundance below $P_{mix}$ any more than did the enriched region model.
Because so little H$_2$S absorption is present, the spectrum of the depleted north polar region is more sensitive than the enriched equatorial region to the NH$_3$ abundance below the mixing pressure, and this quantity is constrained to $1.10_{-0.26}^{+0.45} \times$ Solar (1$\sigma$). Above the mixing pressure, the model is consistent with a factor of $\sim$280 depletion in NH$_3$ in the north polar region compared to the enriched midlatitude region, i.e., an abundance of $\sim$$5\times10^{-7}$.
The retrieved mixing pressure of $55_{-12}^{+9}$ (1$\sigma$) is somewhat deeper than the level at which the NH$_4$SH cloud is expected to form in equilibrium ($\sim$35 bar); however, this value remains within the 95\% (2$\sigma$) confidence interval. Thus it is a reasonable guess that the atmosphere transitions from well-mixed to depleted at the NH$_4$SH cloud.

%

\subsubsection{RT Modeling Discussion}
\label{section_discussion}

The good fit to the data provided by our cloud-physics model (Figure \ref{bestfit}) lends strong support to a Uranian atmosphere dominated by H$_2$S absorption in the upwelling regions, with most or all NH$_3$ removed from the troposphere by the NH$_4$SH cloud. This model, which is similar to that of \citet{depater89, depater91}, permits significant H$_2$S above the NH$_4$SH cloud. The nitrogen-to-sulfur (N/S) ratio does not need to be tuned to very near unity as required by the H$_2$S-inclusive models of \citet{hofstadter92}. The midlatitudes are fit by a solar NH$_3$ abundance below the NH$_4$SH layer, which forms at a pressure of $\sim$35 bar. This is deeper than the midlatitude mixing pressure of $\sim$22 bar determined by \citet{hofstadter90}; however, those authors did not consider the role of H$_2$S absorption on Uranus's radio spectrum. In the downwelling north polar region, our model agrees well with the \citet{hofstadter90} and \citet{hofstadter92} south polar region model: those authors required an NH$_3$ abundance of $\sim$$5\times10^{-7}$ down to $\sim$50 bar pressures (they ignored H$_2$S gas opacity). All three of those requirements are within $1\sigma$ of our results, providing evidence that the north- and south-polar warm spots may arise from the same chemical processes.

The retrieved abundances derived in the previous subsections came from a simplified cloud-physics model that did not take into account the solution cloud at pressures deeper than 50 bars. Thus, to make inferences about the true deep abundances of nitrogen- and sulfur-bearing species, we returned to the full cloud-physics model of \citet{depater91}, which includes prescriptions for the deep water and water-solution clouds from \citet{atreya85}. We tuned the model such that the NH$_3$ and H$_2$S abundances between the solution cloud and the NH$_4$SH cloud matched our retrieved abundances of those species. For this purpose, we used the best-fitting H$_2$S abundance below the NH$_4$SH cloud from the enriched region and the best-fitting NH$_3$ abundance below $P_{mix}$ from the depleted region, since those were the best-constrained values and the two regions are assumed to be well-mixed below the mixing layer. This procedure yielded abundances in the deep troposphere of $1.7_{-0.4}^{+0.7} \times 10^{-4}$ (1.4$_{-0.3}^{+0.6}$$\times$ Solar; 1$\sigma$) for NH$_3$ and $8.7_{-1.5}^{+3.1} \times 10^{-4}$ (37$_{-6}^{+13}$$\times$ Solar; 1$\sigma$) for H$_2$S. 

Using the NH$_3$/H$_2$S ratio as a proxy for the nitrogen/sulfur (N/S) ratio, our results provide a much stronger constraint on Uranus's bulk atmospheric N/S ratio than previous work. We find
N/S = $0.20_{-0.07}^{+0.08}$ (1$\sigma$), in agreement with \citet{depater89, depater91}, who required N/S$<0.2$. This is a very strong selective enrichment in sulfur considering the Solar N/S value of $\sim$5. Our observed sulfur-to-nitrogen ratio can be explained by an ice giant formation scenario in which volatiles were trapped as clathrate hydrates and then swept up by planetesimals \citep[][]{hersant04}. Those authors argue that in the cold, ice-rich conditions of the outer disk, NH$_3$ and H$_2$S would be trapped very efficiently by clathration and accrete onto solid grains, while N$_2$ would remain in the gas phase. At the temperature and pressure of the outer disk, nearly all sulfur is in the form of H$_2$S, and the N$_2$:NH$_3$ ratio is roughly 10:1.  Therefore, Uranus and Neptune should have accreted nearly all available sulfur but only a small fraction of available nitrogen, decreasing the N/S ratio by a factor of $\sim$20, which is close to what we observe.


\section{Conclusions}
\label{section_conclusions}

The millimeter and radio observations presented here provide a unique view of Uranus's troposphere: we provide the first published millimeter-wavelength maps of the planet, as well as the first VLA observations taken during northern summer. Our results are summarized as follows.

\begin{itemize}
	
	\item High-spatial-resolution maps reveal a complex tropospheric circulation pattern, including thin, bright, likely downwelling bands at $\sim$0$^\circ$N, 20$^\circ$N, and 20$^\circ$S in the 2 cm VLA and 2.1/3.1 mm ALMA maps. The identity of the absorber in these bands remains unknown, but variations in the relative humidity of H$_2$S or the methane abundance are good candidates to explain these features. Kinetic temperature variations may also play a role.
	
	\item The north polar region is approximately equal in brightness and extent to the south polar region observed three decades ago, with a radio brightness temperature $\sim$35 K brighter at 2 cm than the midlatitudes. The polar brightening can be observed over a large range of frequencies from 10 cm - 1 mm. Taken together with methane sensitive infrared measurements, this implies a single vertical cell of downwelling air from $\sim$0.1 to 50 bars. Radiative transfer modeling suggests the north polar region is depleted in ammonia to a similar degree as the south polar region observed 30 years ago; both have an NH$_3$ volume mixing ratio of $\sim$$5\times10^{-7}$ above the $\sim$50 bar level and zero H$_2$S opacity.
	
	\item The radio spectrum of the dark upwelling midlatitude regions between $\sim$40$^\circ$S and $\sim$40$^\circ$N is consistent with a model in which all NH$_3$ is removed by the NH$_4$SH cloud at $\sim$35 bar, and constrains the deep H$_2$S abundance to $37_{-6}^{+13} \times$ Solar.
	
	\item The radio spectrum of the downwelling north polar region at latitudes north of $\sim$50$^\circ$N is consistent with strong depletions in both ammonia and hydrogen sulfide from the top of the atmosphere down to a mixing pressure $P_{mix} = 55_{-12}^{+9}$. The strong depletion in H$_2$S in this region permits the deep NH$_3$ abundance to be well constrained, taking a value of $1.4_{-0.3}^{+0.6} \times$ Solar.
	
	\item The deep sulfur-to-nitrogen ratio in Uranus's troposphere is N/S = $0.20_{-0.07}^{+0.08}$ (note protosolar N/S $\sim$ 5), assuming the \citet{atreya85} prescription for the water-solution cloud at pressures deeper than 50 bars and no temperature difference between equator and pole. This is the most stringent constraint on that ratio to date.

	\item The phosphine abundance in the ice giants remains unknown; the observations presented in this work are only sensitive to pressure-broadened phosphine spectral lines at the 2$\sigma$ level in one band (ALMA Band 6; 1.3 mm), and the spectral slope in this band does not support a pressure-broadened phosphine line.

\end{itemize}

This paper lays the groundwork for more detailed studies of Uranus's tropospheric circulation and composition with the upcoming Next-Generation Very Large Array (NGVLA). The NGVLA will perform 10 cm and 20 cm observations at resolutions of $\sim$3 mas and $\sim$10 mas, respectively, permitting a much stronger constraint on the ammonia abundance in the ice giants. Observations from 3 cm to 3 mm at resolutions down to 0.1 mas will resolve the new dark and bright bands discovered in this work well enough to extract robust brightness temperature measurements from those regions; applying radiative transfer models may identify the absorbers responsible for these bands. An outline of the solar system science that will become possible with the ngVLA is given in \citet{depater18}.

\software radiobear, scipy, emcee, CASA, MIRIAD

\acknowledgements

{\large\textit{Acknowledgements:}}

This paper makes use of the following VLA data: VLA/2014-06-232. 
The National Radio Astronomy Observatory (NRAO) is a facility of NSF operated under cooperative agreement by Associated Universities, Inc. VLA data used in this report are available from the NRAO Science Data Archive at https://archive.nrao.edu/archive/advquery.jsp. 

This paper makes use of the following ALMA data: ADS/JAO.ALMA\#2017.1.00855.S. ALMA is a partnership of ESO (representing its member states), NSF (USA) and NINS (Japan), together with NRC (Canada), MOST and ASIAA (Taiwan), and KASI (Republic of Korea), in cooperation with the Republic of Chile. The Joint ALMA Observatory is operated by ESO, AUI/NRAO and NAOJ. ALMA data used in this report are available from the NRAO Science Data Archive at http://almascience.nrao.edu/aq/. 

This research was supported by NASA grant NNX16AK14G through the Solar System Observations (SSO) program to the University of California, Berkeley. 

E. Molter was supported in part by NRAO Student Observing Support grant \#SOSPA6-006.

We thank the staff at the 2018 NRAO Synthesis Imaging Workshop for providing advice on calibration of the ALMA data used in this paper.

\appendix

\setcounter{table}{0}
\renewcommand{\thetable}{A\arabic{table}}
\setcounter{figure}{0}
\renewcommand{\thefigure}{A\arabic{figure}}

\section{Meridional Temperature Gradients}
\label{appendix}

The radio occultation experiment aboard the Voyager 2 spacecraft determined the temperature-pressure profile in Uranus down to roughly 2.7 bar \citep{lindal87}; however, this measurement was made using two occultations between 0$^\circ$ and -10$^\circ$ latitude, so meridional differences were not observed.  The IRIS instrument determined the temperature structure down to $\sim$0.6 bar as a function of latitude, finding temperature differences smaller than 2 K \citep{pearl90}. Similarly small temperature differences have also been observed in the upper troposphere from more recent ground-based studies \citep{roman20}. The latitudinal temperature structure of Uranus has not been directly observed below 1 bar. Most atmospheric models predict that meridional temperature gradients should be small because the orbital period is much shorter than the radiative relaxation time in Uranus's troposphere \citep{wallace83, friedson87, conrath90}; however, recent calculations have cast doubt on the radiative timescales used in those papers \citep{li18}. We must therefore consider the possibility that the observed brightness temperature difference between the midlatitude and polar regions is driven primarily by differences in kinetic temperature, instead of differences in composition as assumed in the main text. The strength of an absorption feature is set by both the abundance profile of the absorbing species and the atmospheric temperature profile. To disentangle temperature and abundance is an underconstrained problem; however, a large midlatitude-to-pole temperature difference between $\sim$0.6 and 50 bar is generally considered unlikely based on several lines of evidence. These have been summarized convincingly by \citet{hofstadter03}, and here we leverage our new data to expand upon the arguments of those authors.

\begin{figure}
	\includegraphics[width = 0.9\textwidth]{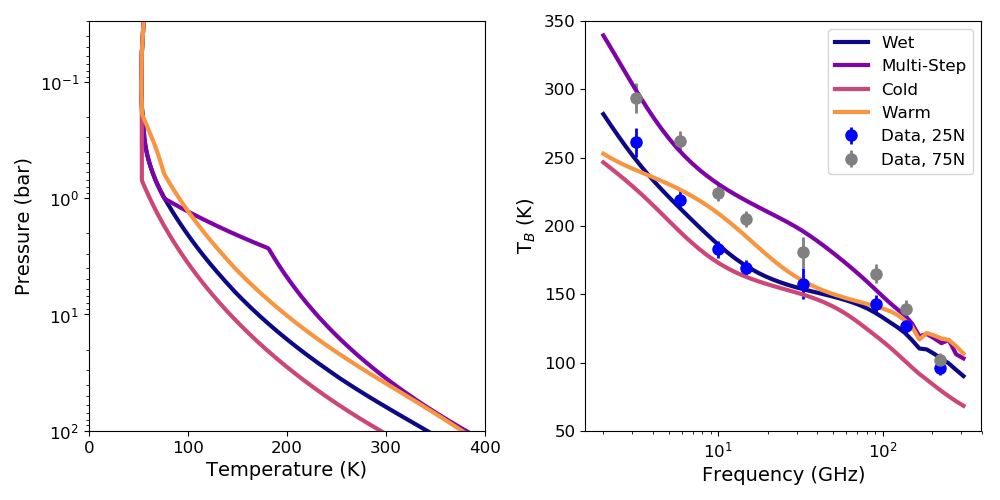}
	\caption{Effect of temperature profile perturbations on the radio/millimeter spectrum of Uranus. Using the same chemical abundances as for our best-fitting model at 25$^\circ$N, we consider a wet adiabat (blue line) and a warmer and cooler adiabat (orange and salmon lines). We also attempt to perturb the temperature profile ad-hoc until a reasonable fit to the north-polar-region data is achieved (purple line).\label{tptest}}
\end{figure}

The radio/millimeter spectra at the midlatitudes and poles cannot both be fit with the same composition unless a highly unphysical temperature-pressure profile is assumed. The latent heat of condensation of H$_2$S (or H$_2$S/NH$_3$ in the NH$_4$SH cloud) at saturation is very small compared to the enthalpy of a parcel of air in Uranus, so the difference between a moist and dry adiabat is very small. However, if the requirement to fit the Voyager data at pressures less than 0.6 bar is relaxed, warmer or cooler adiabats can be considered. Starting with our best-fitting gas abundances and vertical temperature structure in the midlatitude region (Section \ref{section_enriched}), we tried using warmer adiabatic profiles; a sample result is shown in Figure \ref{tptest}. No adiabatic profile can effectively fit the north polar data. To achieve a reasonable fit to the observed north polar spectrum using only kinetic temperature changes requires many unphysical kinks in the temperature structure (Figure \ref{tptest}).  The ``solutions'' we find require a strongly superadiabatic lapse rate near the H$_2$S ice cloud level to bring H$_2$S condensation high in the atmosphere, and a strongly subadiabatic lapse rate deper than that to keep the NH$_4$SH cloud deep enough.  Overall, we find that the spectra, and in particular the millimeter wavelengths, are highly sensitive to the pressure level of the H$_2$S and NH$_4$SH clouds; the kinetic temperature must be tuned very finely to place these at the correct equilibrium level. It is worth noting that a discontinuity in the temperature profile at a condensation layer is possible if a large vertical gradient in atmospheric molecular weight is also present, as has been suggested for both the methane and H$_2$O cloud layers \citep{guillot95, cavalie17, cavalie20}. However, the abundances of NH$_3$ and H$_2$S have a factor of $\gtrsim$25 less effect on the atmospheric molecular weight than those of CH$_4$ and H$_2$O, so this effect is not expected to be important near the NH$_4$SH or H$_2$S cloud layers.

\citet{hofstadter03} considered the complementary situation, in which the bright polar region followed an adiabat and the equatorial region was fit using strongly subadiabatic temperature profiles. This goes contrary to most theory, which generally predicts that the pole should be more stably stratified \citep{briggs80, wallace83, friedson87}, but shall be considered regardless. In this case, the temperature profiles at the midlatitudes and poles were both forced to fit the Voyager data, but allowed to diverge below.  \citet{hofstadter03} note that the temperature differences needed to make this work, which reach 90 K at 20 bar pressure, imply a vertical wind shear of $-100$ m s$^{-1}$ per scale height at 20 bar by the thermal wind equation. However, they stop short of integrating the thermal wind equation vertically to produce a geostrophic wind profile. We have done this, and find that the winds must be supersonic deeper than $\sim$20 bar, which is unphysical. It is worth noting that the thermal wind equation assumes a compositionally constant atmosphere wherein density differences are due to temperature only. Significant meridional gradients in composition could arise on Uranus via gradients in the methane abundance, since methane accounts for up to 5\% of the troposphere by volume and is much heavier than hydrogen and helium \citep{sun91, tollefson18}. However, methane is observed to be depleted in the poles relative to the midlatitudes \citep{karkoschka09, tice13, sromovsky14, sromovsky19, irwin19}, so the implied compositional (and therefore density) gradients would lead to even larger vertical wind shear. The observed thermal wind \citep[e.g., ][]{sromovsky15} also points, again by the thermal wind equation, to a warmer midlatitude region than polar region.

Taken together, these arguments show that composition is the primary driver of the observed brightness temperature differences at millimeter and radio wavelengths. However, the possibility that both temperature and composition change between the midlatitudes and poles cannot be ruled out.  A different assumed polar temperature profile would have a relatively small but perhaps non-negligible effect on the deep NH$_3$ abundance retrieved in this paper.

\clearpage
\bibliography{paper}{}

\end{document}